\documentstyle[12pt,epsf]{article}
\begin{document}
\begin{titlepage}
~\vskip4cm
\begin{center}{\Large\bf Dispersive and chiral symmetry constraints  on the
light meson form factors}
\end{center}
\begin{center}
Irinel Caprini\\
National Institute of Physics and Nuclear Engineering,\\
POB MG 6, Bucharest, R-76900 Romania \end{center}
\vskip1cm
\begin{abstract}
 The 
form factors of the light pseudoscalar mesons are  investigated in a dispersive
formalism  based on hadronic unitarity, analyticity and the OPE expansion of
the QCD Green functions. We propose
generalizations of the original mathematical techniques,  suitable for
including additional  low energy information provided by experiment or Chiral
Perturbation Theory (CHPT).  The simultaneous treatment of  the  electroweak
form factors of the $\pi$ and $K$ mesons
allows us to test the consistency with QCD of a low energy CHPT theorem. 
 By applying the formalism to the pion
electromagnetic form factor, we derive quite strong  
constraints  on the  higher Taylor coefficients at zero momentum, using
information about the phase and the modulus of the form factor along a part of
the unitarity cut. \end{abstract} \end{titlepage} \newpage

\section {Introduction}
Chiral perturbation theory  \cite{GaLe} provides a systematic low energy 
expansion of the QCD Green functions in the sector of
light mesons. Calculations up to two loops were done recently for the 
scattering amplitudes  and the electroweak  form factors in the chiral
$SU(2)\times SU(2)$ limit \cite{GaMe}-\cite{BiCo}. A  full ${\cal
O}(p^6)$  $SU(3)\times SU(3)$ calculation  was also
performed \cite{PoSc} for a specific combination of form factors which does
not involve arbitrary  renormalization constants.
The result allowed to estimate the symmetry breaking
corrections to a low energy theorem proposed by Sirlin \cite{Sirl}, which
generalizes the Ademollo-Gatto theorem \cite{AdGa}.

The fundamental properties of causality and unitarity are important
ingredients in  CHPT.   Dispersion relations for scattering amplitudes
and form factors or their inverses were  used  in performing 
low order  calculations \cite{GaMe}, \cite{BiCo}, or as a method of effective
resummation   of the high order terms  \cite{Truo1}, \cite{Hann}. In the
present paper we shall consider an alternative dispersive method suitable
especially for form factors  \cite{Meim},\cite{Okub}, which uses as input an
information about the modulus along the unitarity cut, leading to constraints
on the values inside the analyticity domain. The
technique was for the first time combined with the  OPE expansion of a QCD
polarization function in \cite{BoMa}.  Recently, this formalism was
successfully applied to the form factors describing the weak semileptonic
decays of heavy mesons \cite{RaTa}-\cite{CaLe}. Quite strong constraints on
the shape of these  form factors near zero recoil have been obtained by
combining the dispersive formalism with the predictions of the heavy quark
effective theory  (HQET) \cite{IsWi}, \cite{Neub}.  The predictive power of the technique
was considerably increased 
\cite{CaMa}, \cite{BoGr1}, \cite{CaLe} by the simultaneous treatment of
several form factors with different unitarity thresholds, connected among them
near zero recoil by HQET.

It is of interest to apply this formalism 
to the form factors of the light pseudoscalar  mesons, for which chiral
symmetry predicts definite correlations for certain kinematical points.
 A first investigation of the weak form factors of
the $K\to \pi l\nu$ decay in this framework was performed in
\cite{BoMa}. Similar techniques were applied in \cite{BuLe} in order to
parametrize the  modulus of  the pion electromagnetic
form factor in the time like  region. In the present paper we perform an
analysis of the electromagnetic and weak form factors of the $\pi$ and $K$
mesons, with emphasis on their low energy expansions. The purpose of the
investigation is to see whether the dispersive approach is useful for testing
the rigorous  predictions of chiral symmetry and for
constraining the free parameters of CHPT. To this end we present
generalizations of the standard technique suitable for incorporating additional
information provided by  experiments or CHPT.

The paper is organized as follows. In the next Section  we review the
dispersive formalism: first we present the form factors of interest
and some of their properties, then we describe the physical input of the method
and  the standard mathematical techniques used for optimally
exploiting  this input. In Section 3 we illustrate the simultaneous
treatment of several form factors with different unitarity branch points by
performing a test of  a low energy theorem of CHPT.  In Section 4 we derive constraints  
on  the Taylor coefficients at zero momentum of the pion electromagnetic form
factor. We first give the simple unitarity bounds obtained from the standard
formalism and then show how they are improved using additional information
on the phase and the modulus of the form factor along a part of the unitarity
cut. In the last Section we present  some conclusions. 
\section {Review of the dispersive formalism}  
 \subsection{Definitions and notations}\label{s1} 
We consider the
electromagnetic form factors of the $\pi$ and $K$ mesons, defined as 
\begin{eqnarray}\label{defe} \langle \pi^+(p')\vert J_\mu^{elm} \vert
\pi^+(p)\rangle= (p+p')_\mu F_{\pi}(q^2)\,\nonumber \\ \langle K^+(p')\vert
J_\mu^{elm} \vert K^+(p)\rangle= (p+p')_\mu F_{K^+}(q^2)\,\nonumber \\ \langle
K^0(p')\vert J_\mu^{elm} \vert K^0(p)\rangle= (p+p')_\mu F_{K^0}(q^2)\,,
\end{eqnarray} and the form factors describing the weak semileptonic decay
$K\to \pi\l \nu$: \begin{eqnarray}\label{defw} \langle \pi^0(p')\vert
J_\mu^{w} \vert K^+(p)\rangle= (p+p')_\mu f^{(+)}_{\pi K}(q^2) +(p-p')_\mu
f^{(-)}_{\pi K}(q^2)\,\nonumber\\ =\left(q_\mu-{q\cdot p\over q^2
}P_\mu\right) f_{\pi k}(q^2) +q_\mu d_{\pi k}(q^2)\,, \end{eqnarray} 
where  $q=p-p'$, $P=p+p'$ and $J_\mu^{elm}$ ($J_\mu^w$) is the
electromagnetic (weak) current:
\begin{equation}\label{current} J_\mu^{elm}={2\over 3}\bar u\gamma_\mu
u-{1\over 3}\bar d\gamma_\mu d- {1\over 3}\bar s\gamma_\mu s\,,\quad
J_\mu^{w}=\bar s\gamma_\mu u \,. \end{equation} The form factors defined in
(\ref{defe}) and (\ref{defw}) are analytic functions  of real type in the
complex plane $t=q^2$ cut along the real axis from the threshold of two
particle production  to infinity. The  cut starts at $t_\pi=4m_{\pi}^2$ for the
electromagnetic form factors  $F_{\pi}(t)\,,\,F_{K^+}(t)$ and $F_{K^0}(t)$,
and  at $t_{\pi K}=(m_\pi+m_K)^2$ in the case of the weak form factors $d_{\pi
K}(t)$ and $f_{\pi K}(t)$.  At $t=0$ the conservation of the electromagnetic
current and the Ademollo-Gatto theorem  \cite{AdGa} give
\begin{eqnarray}\label{norm} F_{\pi}(0)=1\,,\,\,F_{K^+}(0)=1\,,\,\,
F_{K^0}(0)=0\,,\nonumber\\   d_{\pi K}(0)=(m_K^2-m_\pi^2)f_{\pi K}(0)\,,\quad
f_{\pi K}(0)=1\,. \end{eqnarray}  If
we define the function \begin{equation}\label{Delta} \Delta(t)={1\over 2}
F_{\pi}(t)+{1\over 2}F_{K^+}(t)+F_{K^0}(t)-f_{K\pi}(t) \,,
\end{equation}from (\ref{norm}) it follows that \begin{equation}\label{adem}
\Delta(0)=0\,.
\end{equation}
 This relation is valid up to terms quadratic in the
chiral symmetry breaking parameters \cite{AdGa}. The 
Ademollo-Gatto theorem was extended by Sirlin \cite{Sirl} to values
$t\ne 0$ near the origin. Sirlin's theorem requires in particular the vanishing
of the derivatives of the function $\Delta(t)$ at $t=0$:
\begin{equation}\label{sirlin1}
\Delta'(0)=0\,, \,\,\, \Delta''(0)=0\,,.... \end{equation}
In CHPT  the renormalization constants cancel in
the combination of the form factors entering Sirlin's function. A recent
${\cal O}(p^6)$  calculation in full chiral $SU(3)\times SU(3)$
perturbation theory gave the value  \cite{PoSc} 
\begin{equation}\label{post} \Delta'(0)={1\over 6} r^2_S= (0.0033\pm 0.0005)
~{\rm fm}^2\,, \end{equation} where $r^2_S$ is the "charge radius" of the
Sirlin's form factor \cite{PoSc}. This result shows explicitely the
breaking symmetry correction to Sirlin theorem (\ref{sirlin1}). 
\subsection {Unitarity and dispersion inequalities }\label{s2}
Dispersive bounds on the above form factors are obtained by considering the
vacuum polarization tensor :
\begin{equation}\label{polar}
i\int dx e^{iqx}\langle 0\vert T(J_\mu^\dagger (x)J_\nu (0))\vert 0\rangle =
(q_\mu q_\nu-g_{\mu \nu}q^2)\Pi(q^2)+g_{\mu\nu}D(q^2)\,,
\end{equation}  where $J_\mu$  denotes either the electromagnetic
$J_\mu^{elm}$  or the weak  $J_\mu^{w}$  current (the
function $D(q^2)$ vanishes in the electromagnetic case). From
the asymptotic behaviour of QCD it follows  that the derivative  $\Pi' (q^2)$ 
of the amplitude  $\Pi (q^2)$ satisfies the dispersion relation 
\begin{equation}\label{disprel} \Pi'(q^2)={1\over
\pi}\int\limits_0^\infty{{\rm Im}\Pi (t+i\epsilon) \over (t-q^2)^2}{\rm d}t\,,
\end{equation}   with the spectral function  ${\rm Im}\Pi(t+i\epsilon)$ given
by hadronic unitarity. Using the definitions (\ref{defe}) of the 
electromagnetic form factors and taking into account the positivity  ${\rm
Im}\Pi(t+i\epsilon)\ge 0$, we obtain  the inequality:
\begin{eqnarray}\label{unitelm}  
{\rm Im}\Pi_{elm}(t+i\epsilon) \geq  
{1\over 48\pi}
 \left(1-{t_\pi\over t}\right)^{3/2} 
\vert F_{\pi}(t)\vert^2\theta(t-t_\pi)\nonumber\\
+ {1\over 48\pi}\left(1-{t_K\over t}\right)^{3/2} 
\left[\vert F_{K^+}(t)\vert^2+\vert F_{K^0}(t)\vert^2\right]\theta(t-t_K)\,,
\end{eqnarray}  where $t_\pi=4m_\pi^2$ and 
$t_K=4m_K^2$ are unitarity branch points.   By inserting the
inequality (\ref{unitelm})
in the  dispersion relation  (\ref{disprel}) for an euclidian point
$q^2=-Q^2<0$ we obtain:  \begin{eqnarray}\label{condelm}  
&&\Pi'_{elm}(-Q^2)\geq  
{1\over 48\pi^2}\int\limits_{t_\pi}^\infty{{\rm d}t \over (t+Q^2)^2} 
\left(1-{4m_\pi^2\over t}\right)^{3/2}  
\vert F_{\pi}(t) \vert ^2 \nonumber \\ 
&+&{1\over 48\pi^2}\int\limits_{t_K}^\infty{{\rm d}t \over (t+Q^2)^2} 
\left(1-{t_K\over t}\right)^{3/2}  
\left[\vert F_{K^+}(t) \vert ^2 +\vert F_{K^0}(t)\vert^ 2 \right] \,. 
\end{eqnarray} 
A relation similar to (\ref{condelm}) can be written for the polarization
function $\Pi_{w}(Q^2)$ of the weak current. Using the definition (\ref{defw})
we obtain from unitarity and positivity 
\begin{eqnarray}\label{unitw}   {\rm Im}\Pi_{w}(t+i\epsilon) \geq  
{\eta\over 48\pi}\left(1-{t_{\pi K}\over
t}\right)^{1/2}  \left(1-{t_{\pi K}^{(-)}\over t}\right)^{1/2}\times\nonumber\\
\left[\left(1-{t_{\pi K}\over t}\right) \left(1-{t_{\pi K}^{(-)}\over t}\right)
 \vert f_{\pi K}\vert^2+ 
  {\vert d_{\pi K}\vert^2\over t^2}\right] 
\theta(t-t_{\pi K})\,,
\end{eqnarray} 
where $t_{\pi K}=(m_K+m_\pi)^2$,  $t_{\pi K}^{(-)}=(m_K-m_\pi)^2$
and  $\eta= {3\over 2}$  is an isospin factor \cite{BoMa}.   By inserting 
(\ref{unitw}) in a dispersion relation of the form (\ref{disprel}) we obtain  
\begin{eqnarray}\label{condw}   &&\Pi'_{w}(-Q^2)\geq   {\eta\over
48\pi^2}\int\limits_{t_{\pi K}}^\infty{{\rm d}t\over (t+Q^2)^2} 
\left(1-{t_{\pi K}\over t}\right)^{1/2} 
\left(1-{t_{\pi K}^{(-)}\over t}\right)^{1/2} \nonumber \\
&&\times \left[\left(1-{t_{\pi K}\over t}\right)
\left(1-{t_{\pi K}^{(-)}\over t}\right) 
 \vert f_{\pi K}(t)\vert^2+ 
  {\vert d_{\pi K}(t)\vert^2\over t^2}\right] \,. 
\end{eqnarray} 
It is interesting to note that the inequalities (\ref{condelm}) and
(\ref{condw}) can be added, leading to a single inequality  
  \begin{eqnarray}\label{condit}  
&&\Pi'(-Q^2)\geq   
{1\over 48\pi^2}\int\limits_{t_\pi}^\infty{{\rm d}t \over (t+Q^2)^2} 
\left(1-{4m_\pi^2\over t}\right)^{3/2}  
\vert F_{\pi^+}(t) \vert ^2 \nonumber \\ 
&+&{1\over 48\pi^2}\int\limits_{t_K}^\infty{{\rm d}t \over (t+Q^2)^2} 
\left(1-{t_K\over t}\right)^{3/2}  
\left[\vert F_{K^+}(t) \vert ^2 +\vert F_{K^0}(t)\vert^ 2 \right] \nonumber\\
&+& 
{\eta\over 48\pi^2}\int\limits_{t_{\pi K}}^\infty{{\rm d}t\over (t+Q^2)^2} 
\left(1-{t_{\pi K}\over t}\right)^{1/2} 
\left(1-{t_{\pi K}^{(-)}\over t}\right)^{1/2} \nonumber \\
&&\times \left[\left(1-{t_{\pi K}\over t}\right)
\left(1-{t_{\pi K}^{(-)}\over t}\right) 
 \vert f_{\pi K}(t)\vert^2+ 
  {\vert d_{\pi K}(t)\vert^2\over t^2}\right] \,,
\end{eqnarray} where
\begin{equation}\label{pi0} \Pi'(-Q^2)=\Pi_{elm}'(-Q^2)+\Pi_{w}'(-Q^2)\,.
\end{equation}
In the euclidian region $Q^2>0$, the functions  $\Pi'_{elm}(-Q^2)$ and 
$\Pi'_w(-Q^2)$ can be calculated by applying renormalization group improved
perturbative QCD, with nonperturbative
corrections included by means of operator product expansions
 (OPE). We 
used the expressions given in  \cite{BrNa}:
\begin{eqnarray}\label{pi01} &&\Pi'(-Q^2)= \nonumber\\ &&=\left({1\over
6}+{1\over 4}\right) {1\over \pi^2 Q^2}\left[1+ {\alpha_s(-Q^2)\over \pi}+F_3
\left({\alpha_s(-Q^2)\over \pi}\right)^2 +\left(F_4+{\beta_1^2\pi^2\over
12}\right) \left({\alpha_s(-Q^2)\over \pi}\right)^3\right] \nonumber\\
&&-\left({1\over 6}+{3\over 4}\right){1\over \pi^2 Q^4}
m_s^2(-Q^2)+\left({1\over 18}+{1\over 12}\right){1\over
Q^6} \langle{\alpha_s\over
\pi}\bar G G\rangle + \left({10\over 9}+1\right) {1\over Q^6} m_q\langle
\bar\psi\psi\rangle\,, \end{eqnarray}  corresponding to  $n_f=3$ flavours. The
first and the second numerical constant in front of each term indicate  the 
separate contribution of  $\Pi_{elm}'(-Q^2)$ and $\Pi_{w}'(-Q^2)$,
respectively.  To evaluate these quantities  we used the  two-loop
expressions of the running coupling $\alpha_s(-Q^2)$ and the running mass 
$m_s(-Q^2)$ of the $s$-quark \cite{BrNa}, the
perturbative parameters  $F_3=1.6398\,,\, F_4=-10.2839\,, \,\beta_1=-9/2$
\cite{BrNa} and  the condensates $m_q\langle \bar\psi\psi\rangle = (0.200\,
{\rm GeV})^4$ and $\langle \alpha_s\bar G G\rangle/\pi=(0.45 \,{\rm GeV})^4$
\cite{SVZ}. As concerns the value of the euclidian point $Q^2$ at which we
calculate the polarization amplitude,  it must be  on one hand high enough to
ensure the validity of the OPE expansion, and on the other hand small enough to
provide a strong constraint on the form factors through  the dispersion
relation (\ref{disprel}). In fact, the results turn out to be rather stable
when $Q^2$ is varied in the range $2\, {\rm GeV}^2\leq Q^2\leq 6\, {\rm
GeV}^2$. The results reported in this work are obtained with the choice  
$Q^2=2\, {\rm GeV}^2$, for which $\Pi_{elm}'(-Q^2)=0.009546 \,{\rm GeV}^{-2}$
and $\Pi_{w}'(-Q^2)=0.0133 \,{\rm GeV}^{-2}$.  With the l.h.s. of the relations
(\ref{condelm}), (\ref{condw}) or (\ref{condit})   known, these inequalities
are  integral  conditions for the sum of the moduli squared of the
corresponding form factors.

It is useful to point out that an
alternative integral condition for the electromagnetic form factors can be
obtained by using  as input, instead of OPE in the euclidian region, a lower
bound on the hadronic muon anomaly.  This method was first proposed in
\cite{GoRa}, and starts from the the vacuum polarization contribution of the
hadronic part of the muon anomaly \begin{equation}\label{amu}
a^{(h)}_\mu={1\over \pi}\int\limits_0^\infty {{\rm d}t\over t}  {\rm Im}
\Pi(t+i\epsilon) {\cal K}(t)\,, \end{equation} where
\begin{equation}\label{K}
{\cal K}(t)={\alpha\over \pi}
\int\limits_0^1{z^2(1-z)\over z^2+(1-z)t/ m_\mu^2}{\rm d}z \,.
\end{equation} By introducing  (\ref{amu}) in the unitarity
relation (\ref{unitelm}) we obtain the inequality
\begin{eqnarray}\label{condit1} 
a_\mu^{(h)}\geq 
{1\over 48\pi^2}\int\limits_{t_\pi}^\infty{K(t)\over t}
\left(1-{4m_\pi^2\over t}\right)^{3/2} 
\vert F_{\pi^+}(t) \vert ^2 {\rm d}t\nonumber \\
+{1\over 48\pi^2}\int\limits_{t_K}^\infty {K(t)\over t}
\left(1-{4m_K^2\over t}\right)^{3/2} 
[\vert F_{K^+}(t) \vert ^2 +\vert F_{K^0}(t) \vert ^2 ]{\rm d}t\,,
\end{eqnarray} which restricts the
electromagnetic form factors along the unitarity cut. In the
calculations we adopted  the  conservative lower bound $a^{(h)}_\mu \ge 
7.5\times 10^{-8}$ \cite{AlDa},
based on the experimental value, 
the weak radiative corrections and an estimate of the hadronic box diagram.
\subsection{Standard mathematical techniques}\label{s3} In this subsection we
briefly review the standard techniques for exploiting in the optimal way  the
conditions (\ref{condelm}),  (\ref{condw}), (\ref{condit})  or (\ref{condit1})
in order to obtain constraints on  the form factors and their derivatives
inside the analyticity domain.  It is useful to introduce a compact notation,
defining  the form factors $F_i(t)\,,\,\, i=1,..\,,5$  as
\begin{eqnarray}\label{notation}
F_1(t)=F_\pi(t)\,,\,\,F_2(t)=F_{K^+}(t)\,,\,\,F_3(t)=F_{K^0}(t)\,\nonumber\\
F_4(t)=f_{\pi K}(t)\,,\,\, F_5(t)=d_{\pi K}(t)\,, \end{eqnarray} and denoting
by $t_i$ the unitarity branch points: \begin{eqnarray}\label{ti}
t_1=t_\pi\,,\,\,t_2=t_K\,,\,\,t_3=t_K\,,\nonumber\\
t_4=t_{\pi K}\,,\,\, t_5=t_{\pi K}\,.
\end{eqnarray}
 We now apply to each integral with the lower limit $t_i$ 
appearing in  the conditions (\ref{condelm}), (\ref{condw}),
(\ref{condit}) or (\ref{condit1}) the following conformal mapping
\begin{equation}\label{mapp} z(t)={\sqrt {t_i-t}-\sqrt{t_i}\over \sqrt
{t_i-t}+\sqrt{t_i}}\,. \end{equation}
By this transformation  the  complex $t$ plane is mapped  onto the  unit
disk in the complex  plane  $z$, such that $z(0)=0$ and the  unitarity cut
$t\ge t_i$ becomes the boundary $\vert z \vert = 1$. 
For simplicity we denote $F_i(z)= F_i(t(z))$, where $t(z)$
is the inverse of the mapping $z(t)$ .
Then  the conditions (\ref{condelm}),  (\ref{condw}),
(\ref{condit}) or (\ref{condit1}) can be written in the canonical form: 
\begin{equation}\label{l2norm}
{1\over 2\pi} \sum\limits_{i=1}^I\int\limits_0^{2\pi}
 \vert w_i(\zeta) F_i(\zeta)
\vert ^2 {\rm d}\theta\leq 1\,,\quad\zeta={\rm exp}(i\theta)\,,
\end{equation} where $I=3$ or $I=5$. In  Eq.(\ref{l2norm}) $w_i(z)$ 
 are analytic functions without zeros inside the unit
disk, their modulus square on the boundary are equal to the weight  functions
appearing in front of the form factors,
multiplied by the Jacobian of the conformal mapping (\ref{mapp}).
In mathematical books these functions are called "outer
functions"  and are defined in terms of their modulus
on the boundary as \cite{Dure}
\begin{equation}\label{wouter} 
 w_i(z)={\rm exp}\left[{1\over 2\pi}\int\limits_0^{2\pi} 
 {\zeta+z\over \zeta-z} 
 \ln |w_i(\zeta)| {\rm d}\theta\right]\,, \quad\zeta={\rm exp}(i\theta)\,.
\end{equation} 
 In our case the functions $w_i(z)$ have simple
explicit expressions  \cite{BoMa}-\cite{CaLe}. 
For the global condition (\ref{condit}) which involves all the form factors,
the outer functions $w_i(z)$ are
\begin{equation}\label{w123}  w_i(z)={(1-d_i)^2\over 16}\,\sqrt{{1\over 6\pi
t_i \Pi'(-Q^2)}}\, {(1+z)^2 \sqrt{1-z}\over (1-zd_i)^2}\,,\quad
i=1,2,3\,, \end{equation} 
\begin{eqnarray}\label{w45}  w_4 (z)&=&
{(1-d_4)^2\over 32 (1-z_-)^{3/2}}\,\sqrt{{\eta_4\over 6\pi t_4
\Pi'(-Q^2)}}\,{(1+z)^2 \sqrt{1-z}(1-zz_-)^{3/4}\over (1-zd_4)^2}\,,\nonumber\\
 w_5 (z)&=&{(1-d_5)^2\over 32 t_5 (1-z_-)^{5/2}}\,\sqrt{{\eta_5\over 2\pi t_5
\Pi'(-Q^2)}}\,{(1+z) (1-z)^{5/2}\over (1-zz_-)^{1/4}(1-zd_5)^2}\,,
\end{eqnarray} where 
\begin{equation}\label{di}
d_i={\sqrt{t_i+Q^2}-\sqrt{t_i}
\over\sqrt{t_i+Q^2}+\sqrt{t_i}}\,,\quad i=1,..\,,5\,,
\end{equation} and \begin{equation}\label{z_-}
z_-= {\sqrt{t_{\pi K}-t_{\pi K}^{(-)}}-\sqrt{t_{\pi K}}
\over\sqrt{t_{\pi K}-t_{\pi K}^{(-)}}+\sqrt{t_{\pi K}}}\,.
\end{equation}
For the alternative condition 
(\ref{condit1})  the functions 
$w_i(z)$  are replaced by : 
 \begin{equation}\label{wamu} w_i (z)={1\over
16}\sqrt{{1\over 6\pi t_i a^{(h)}_\mu}}\,(1+z)^2 \sqrt{1-z}~w_{\cal
K}(z)\,,\quad i=1,2,3\,, \end{equation}  where
 $w_{\cal K}(z)$ is an outer function defined as in
(\ref{wouter}) in terms of its modulus  $|w_{\cal K}(\zeta)|=\sqrt{{\cal
K}(\zeta)}$ on the boundary. 

The inequality (\ref{l2norm}) is a standard  boundary condition in $L^2$
norm  for the functions $F_i(z)$ \cite{Dure}, from which one obtains
constraints on the size and shape of these functions at interior points.
The results are expected to be stronger if one  keeps  in  (\ref{l2norm}) the
contributions of all the functions $F_i(z)$  and exploits in addition the
 correlations among them provided by symmetries at some kinematical
points.  This technique proved to be very useful in
the case of the weak form factors of heavy mesons, using 
HQET near zero recoil. In principle, a difficulty in applying the formalism
to the excited states is the presence of the unphysical cuts below the
unitarity branch points. In the case of heavy mesons, the unphysical
cuts are well approximated  by a
few narrow resonances of known masses, which can be treated by adequate
techniques with no assumption about the residua \cite{RaTa}-\cite{CaLe}. 
As we shall see in the next Section, the situation is more complicated in the
case of the light mesons.
\section{Test of a CHPT low energy theorem} In applying the relation
(\ref{l2norm}) we recall that the  functions $F_1, F_4$ and $F_5$,   {\it i.e.}
the electromagnetic form factor of the pion and the weak form factors are
analytic below their unitarity branch points, while the kaon
electromagnetic form factors,  denoted here as $F_2$ and $F_3$,  have an
unphysical cut along the region $t_\pi<t<t_K$,  below the
unitarity threshold  of the  $K \bar K$ production. In the variable $z$ this
cut  is placed inside the unit disk $\vert z\vert < 1$, along the segment
$-1<z< z_0$, where $z_0$ denotes the image of the two pion production
threshold by the  conformal transformation (\ref{mapp}) for the kaon form
factors: \begin{equation}\label{z0} z_0={\sqrt {t_K-t_\pi}-\sqrt{t_K} \over
\sqrt {t_K-t_\pi}+\sqrt{t_K}}\,. \end{equation}  The dispersive
formalism can be applied to this case only if some assumptions about the
unphysical cut are adopted. For instance, if  the phase of the kaon form
factors  along the unphysical cut is supposed to be known, the problem can be
treated by means of a suitable modification of the standard techniques.    
Then one can use the inequalities   (\ref{condelm}) or (\ref{condit}) to
correlate the low energy expansions of the electromagnetic  and weak form
factors or to test some predictions of chiral symmetry.

In this Section we will illustrate  the procedure by testing
the consistency of the relations (\ref{sirlin1}) proposed
by Sirlin \cite{Sirl}, with the requirements of unitarity and perturbative QCD 
introduced in the condition (\ref{condit}). We adopt for the phase of
the kaon electromagnetic form factors a  model similar to the one proposed in
\cite{GuPi} for the pion form factor, and generalized to include a narrow
$\omega$  resonance, with the SU(3) relation  
$C_\rho=3C_\omega$ between the residua. We assumed therefore that the phases of
the form factors $F_{K^+}(t+i\epsilon)$  and $F_{K^0}(t+i\epsilon)$ along the
unphysical region are  \begin{equation}\label{delta+}
\delta_{F_{K^+}}(t)={\rm arctg} \left[{3 m_\rho \Gamma_\rho(t) \over
4(m_\rho^2-t)}\right]\,,\quad t_\pi<t<t_K\,\end{equation} and 
\begin{equation}\label{delta0}  \delta_{F_{K^0}}(t)={\rm arctg}\left[ {3
m_\rho \Gamma_\rho(t)  \over 2(m_\rho^2-t)}\right]\,,\quad t_\pi<t<t_K\,.
\end{equation}  We used the $\rho$ width $\Gamma_\rho
(t)$ given by the resonance chiral effective theory \cite{EcGa1},
\cite{EcGa2}:  
 \begin{equation}\label{gamaro}  \Gamma_\rho(t)={m_\rho
t\over 96 \pi f_\pi^2}\,\left(1-{4m_\pi^2\over t}\right)^{3/2}\,,
\end{equation}  where
$f_\pi=93.1{\rm MeV}$ is the pion decay constant. 

 We shall use now  the information on the
phase in order to remove the unphysical cut of the kaon form factors. To this
end we shall use a so-called Omn\`es  function \cite{Omne}, which exactly
compensates  the known phase of the form factors along this part of the cut,
leaving us with functions analytic below the unitarity threshold.  In what
follows we describe briefly the method.  Since we treat  all the form factors
simultaneously it is convenient to introduce a collective notation, defining
\begin{eqnarray}\label{O_i} O_i(t)=1\,,\quad i=1, 4, 5\,,\nonumber\\
O_i(t)={\rm exp}\left[{t\over \pi}\int\limits_{t_\pi}^{\infty}{\delta_i(t')
\over t'(t'-t)}{\rm d}t'\right]\,,\quad i=2,3\,, \end{eqnarray} where  the
functions $\delta_i(t)$ are defined as 
$\delta_2(t)=\delta_{F_{K^+}}(t)\,,\,\delta_3(t)=\delta_{F_{K^0}}(t)$ for $
t_\pi\le t \le t_K\,$,  and are extended  for $t>t_K$ as Lipschitz continuous
functions \cite{Dure}. Of course, there is a large arbitrariness in such an
extension, but the results are independent of the particular values of
$\delta_i(t)$ for $t>t_K$. The reason is that, being analytic and  without
zeros  in the $t$-plane cut along $(t_\pi, \infty)$,  and with regular  values
on the boundary, $O_2(t)$ and  $O_3(t)$ are outer functions, 
so by multiplying a class of functions with them, the class is not changed 
\cite{Dure}. By construction, the phase of $O_i(t+i\epsilon)$ coincides with
$\delta_i(t)$ .   We also use the notation $O_i(z)$ to denote these functions
in terms  of the variable $z$ defined in (\ref{mapp}). The functions $O_i(z)$
are analytic in $|z|<1$, excepted for a cut along the segment $(-1, z_0)$,
where the phase of  $O_i(z-i\epsilon)$ is equal to  $\delta_i(t(z))$ (we
recall that the conformal transformation (\ref{mapp}) maps the upper half
$t$-plane onto the lower semidisk in the $z$-plane). We define the set of
functions $f_i(z)$ as  \begin{equation}\label{f_i} f_i(z)=F_i(z)
[O_i(z)]^{-1}\,,\quad i=1,..\,,\,5\,, \end{equation}  where $F_i(z)$ are the
form factors defined in (\ref{notation}). From (\ref{f_i}) it is easy to see
that the complex phases of  the form factors $F_2(z)$ and  $F_3(z)$ along the
cut $(-1\,,\, z_0)$ are  compensated by the phases of the Omn\`es functions
$O_i(z)$. Therefore, all the functions   $f_i(z)$ are analytic of real type
inside the unit disk $|z|<1$.  By introducing (\ref{f_i}) in the $L^2$ norm
condition (\ref{l2norm}) we have \begin{eqnarray}\label{l2norm1}
{1\over 2\pi} \sum\limits_{i=1}^5\int\limits_0^{2\pi}  \vert w_i(\zeta)
O_i(\zeta) f_i(\zeta) \vert ^2 {\rm d}\theta = \nonumber\\ {1\over 2\pi}
\sum\limits_{i=1}^5\int\limits_0^{2\pi}  \vert w_i(\zeta) \omega_i(\zeta)
f_i(\zeta) \vert ^2 {\rm d}\theta  \leq 1\,, \quad\zeta={\rm exp}(i\theta)\,,
\end{eqnarray} where $w_i(z)$ are the outer functions defined in (\ref{w123})
and (\ref{w45}), and $\omega_i(z)$ are additional outer functions which 
satisfy the relations \begin{equation}\label{omega_i} \vert \omega_i(\zeta)
\vert = \vert O_i(\zeta)\vert \,,\quad\zeta={\rm exp}(i\theta)\,,\quad
i=1,...5\,. \end{equation} They can be calculated in terms of their modulus on
the boundary using  the standard formula \cite{Dure} given in (\ref{wouter}), 
which is  equivalent to  \begin{equation}\label{omega1}  \omega_i(t)
={\rm exp} \left[{\sqrt{t_K-t}\over\pi}\int\limits_{t_K}^\infty  {\ln \vert
O_i(t')\vert \over \sqrt{t'-t_K} (t'-t)}{\rm d}t'\right]\,,\quad i=1,....5\,.
\end{equation} 
It is convenient to introduce the new functions
$g_i(z)$ as \begin{equation}\label{g_i} g_i(z) =w_i(z) \omega_i(z)
f_i(z)\,=\,\Omega_i(z) F_i(z)\,,\quad i=1,..\,,5\,, \end{equation} where 
\begin{equation}\label{Omega_i} \Omega_i(z)= w_i(z) \omega_i(z)
[O_i(z)]^{-1}\,. \end{equation} Now, in terms of $g_i(z)$ the
inequality (\ref{l2norm1}) takes the canonical form
\begin{equation}\label{l2norm2} {1\over 2\pi}
\sum\limits_{i=1}^5\int\limits_0^{2\pi}  \vert  g_i(\zeta) \vert ^2 {\rm
d}\theta\leq 1\,,\quad \zeta={\rm exp}(i\theta)\,. \end{equation} By
construction the functions $g_i(z)$ are analytic inside the disk $|z|<1$ and
can be expanded  as power series  \begin{equation}\label{g_iseries}
g_i(z)=\sum\limits_{n=0}^\infty c_{n,i} z^n\,, \quad i=1,..5\,, \end{equation}
with real coefficients $c_{n,i}=c^*_{n,i}$.
By inserting these expansions in (\ref{l2norm2}) we obtain
\begin{equation}\label{l2norm3}
 \sum\limits_{i=1}^5 \sum\limits_{n=0}^\infty  c_{n,i}^2 \leq 1\,.
\end{equation}  As we discussed in Section 2, the generalization of the
Ademollo-Gatto  theorem proposed by Sirlin requires the vanishing of the first
derivatives at $t=0$  of the function $\Delta(t)$ defined in (\ref{Delta}), in
the exact chiral limit. In order to test the consistency of this condition,
let us put \begin{equation}\label{Deltak}  
\Delta^{(k)}(0)=s_k\,,\quad k=1,....K\,,  \end{equation}                                                              
where the derivatives  are with respect to $t$, and $s_k$ are  given numbers of
the order of magnitude of chiral symmetry breaking \cite{PoSc}. 
Using (\ref{g_i}) we write  $\Delta(t)$ as
\begin{equation}\label{Delta1} \Delta(t)=\sum\limits_{i=1}^5 a_i \,[\Omega_i
(z)]^{-1}\, g_i(z)\,, \end{equation}
where 
\begin{equation}\label{a_i}
a_1= {1\over 2}\,,\,\, a_2={1\over 2}\,,\,\, a_3=1\,,\,\,  a_4=-1\,,\,\,
 a_5=0\,.
\end{equation} 
Therefore the derivatives entering (\ref{Deltak}) become
\begin{equation}\label{Deltak1}  
\Delta^{(k)}(0)=\sum\limits_{i=1}^5 a_i\sum\limits_{n=0}^k\,C_k^n\left[{{\rm
d}^n [\Omega_i(z)]^{-1} \over {\rm d}t^{n}}\right]_{t=0}\, \left[{{\rm d}^{k-n}
g_i(z) \over {\rm d}t^{k-n}}\right]_{t=0} \,,  \end{equation} and finally 
 the conditions (\ref{Deltak}) can be written as 
\begin{equation}\label{Deltak2}   \sum\limits_{i=1}^5
\sum\limits_{n=0}^k b_{kn}^{(i)}c_{n,i}=s_k\,,\quad k=1,.. ,K\,,
\end{equation} where
the numbers $b_{kn}^{(i)}$ can be obtained  in a
straightforward way using  Eq. (\ref{Deltak1}), the definition (\ref{Omega_i})
of the functions $\Omega_i(z)$ and the connection (\ref{mapp})
between the variables $t$ and $z$. Of course, for each $k$ only the 
Taylor coefficients  $c_{n,i}$ with $n\le k$ contribute to $\Delta^{(k)}(0)$.

Our objective is to test the consistency of the relations (\ref{Deltak})
with the inequality (\ref{l2norm}) provided by the dispersive formalism. The
problem can be easily solved, since we expressed both these conditions in
terms of the Taylor coefficients $c_{n,i}$, transforming in
this way a functional problem into an algebraic one. 
We notice   that the first coefficients 
 \begin{equation}\label{c_0i}
c_{0,i}= g_i(0)=\Omega_i(0) F_i(0)\,,\quad i=1,..\,,5\,
\end{equation}
can be computed  using the relations  (\ref{norm}) and (\ref{Omega_i}). 
The remaining coefficients $c_{n,i}$, $n\ge 1$ are free. The relations
(\ref{Deltak2}) can be viewed as a set of $N$ linear constraints for these
coefficients, which satisfy in addition the quadratic condition
(\ref{l2norm3}). A simple way to test the consistency of these relations for a
given set of numbers $s_k$, $k=1,..\,,\,K$ is to define the quantity
\begin{equation}\label{mu_0}
\mu_0^2= \min_{c_{n,i}} \sum\limits_{i=1}^5
\sum\limits_{n=0}^\infty  c_{n,i}^2\,, \end{equation}  
where the minimization  is with respect to the coefficients
$c_{n,i}$, $n\ge 1\,,\, i=1,....5$ which satisfy the set of linear conditions
(\ref{Deltak2}).
 It is clear that,
if the minimal value $\mu_0^2$ is  greater than one,  the conditions 
(\ref{l2norm3}) and  (\ref{Deltak2}) can not be satisfied  simultaneously,
while if $\mu_0^2$ is less than one  the
conditions are satisfied by at least one
set of coefficients $c_{n,i}$.  A more detailed reasoning along these lines 
\cite{CaDi} shows that the inequality \begin{equation}\label{consis}  \mu_0^2
\leq 1\,, \end{equation}  is a necessary and sufficient condition for the
consistency of the relations  (\ref{l2norm3}) and  (\ref{Deltak2}).

In order to solve the minimization problem (\ref{mu_0})
with the constraints (\ref{Deltak2}) we apply the technique of Lagrange
multipliers. The Lagrangean of the problem is:
\begin{equation}\label{Lagrange}
{\cal L}= \sum\limits_{i=1}^5  c_{0,i}^2
+\sum\limits_{i=1}^5 \sum\limits_{n=1}^\infty  c_{n,i}^2
-2\sum\limits_{k=1}^K\lambda_k \left[ \sum\limits_{i=1}^5 b_{k0}^{(i)}c_{0,i}+
\sum\limits_{i=1}^5 \sum\limits_{n=1}^K b_{kn}^{(i)}c_{n,i}-s_k\right]\,,
\end{equation} 
where $\lambda_k$, $k=1,..K$ are Lagrange multipliers.  In (\ref{Lagrange})  we
indicated separately the coefficients $c_{0,i}$ which are known. Moreover, for
convenience we extended the sum over $n$ in the last term
up to $n=K$, setting $b_{kn}^{(i)}=0$
for $n>k$. The Lagrangean  ${\cal L}$  is  a convex function of the
coefficients $c_{n,i}$, its  minimal value being given by the equations
\begin{equation}\label{minim} {\partial{\cal L}\over \partial
c_{n,i}}=0\,,\quad n=1, ...\infty\,;\,\,\, i=1,..5\,. \end{equation}  The
solution of these equations is \begin{eqnarray}\label{c_n}
c_{n,i}&=&\sum\limits_{k-1}^K\lambda_k b_{kn}^{(i)}\,, \quad n\leq K\,, \,\,\,
i=1,..5\,,\nonumber\\ c_{n,i}&=&0\,,\quad n\ge K+1, \,\,\,
i=1,..5\,.
\end{eqnarray} 
By introducing this solution in the conditions (\ref{Deltak}) we obtain the
following system of equations for the Lagrange multipliers $\lambda_k$:
\begin{equation}\label{system}
\sum\limits_{m=1}^K U_{km} \lambda_m=\tilde s_k\,,\quad k=1,...K\,,
\end{equation} 
where the matrix $U$ is defined as
\begin{equation}\label{U}
U_{km} =\sum\limits_{i=1}^5\sum\limits_{j=1}^K b_{kj}^{(i)} b_{mj}^{(i)}\,,
\quad k,m=1,...K\,,
\end{equation} 
and  the numbers $\tilde s_k$ are related to the derivatives of the Sirlin
function  $s_k$ defined in (\ref{Deltak}) by 
\begin{equation}\label{tildes_k}
\tilde s_k=s_k-\sum\limits_{i=1}^5 b_{k0}^{(i)} c_{0,i}\,,\quad k=1,...K\,.
\end{equation} 
The system (\ref{system}) has the solution  
\begin{equation}\label{lambdak}
 \lambda_k= \sum\limits_{m=1}^K (U^{-1})_{km} \tilde s_m\,,\quad k=1,...K\,,
\end{equation} 
where $U^{-1}$ is the inverse of the matrix $U$. By introducing these
expressions in (\ref{c_n}) we obtain the optimal 
coefficients $c_{n,i}$ for $1\le n\le K$ as
\begin{equation}\label{solution1}
c_{n,i}=\sum\limits_{m=1}^K 
\tilde s_m \sum\limits_{k=1}^K (U^{-1})_{km} b_{kn}^{(i)}\,,\quad
n=1,...K\,;\, i=1,...5\,. \end{equation} 
 With these coefficients for $n\leq K$ and $c_{n,i}=0$ for $n\ge K+1$ according
to (\ref{c_n})  we obtain after a straightforward calculation  the following
compact form the solution  $\mu_0^2$ of the minimization problem (\ref{mu_0}):
 \begin{equation}\label{minnorm1} \mu_0^2= \sum\limits_{i=1}^5 
c_{0,i}^2+\sum\limits_{j,k=1}^K (U^{-1}) _{jk} \tilde s_j \tilde s_k\,,
\end{equation}     
and the consistency condition (\ref{consis}) becomes 
\begin{equation}\label{consis1} 
\sum\limits_{i=1}^5 
 c_{0,i}^2+\sum\limits_{j,k=1}^K (U^{-1}) _{jk} \tilde s_j \tilde s_k\leq
1\,. \end{equation}  
We recall that in this inequality the coefficients $c_{0,i}$ are known
according to (\ref{c_0i}) from the normalization of the form factors at $t=0$,
and the matrix $U$ can be explicitely computed using (\ref{U}) and the
numbers $b_{kn}^{(i)}$ appearing in (\ref{Deltak2}). We mention that these
 numbers include the
physical input on unitarity and QCD ,  contained in the outer functions
$w_i(z)$,  and the phase of the kaon form factors along the unphysical region
entering the outer functions $O_i(t)$ and $\omega_i(t)$. As
for the quantities $\tilde s_k$, they  are related to the derivatives $s_k$ of
the Sirlin function, according to (\ref{tildes_k}). With this input the
inequality (\ref{consis1}) is a constraint for the derivatives $s_k$ of the
Sirlin function $\Delta(t)$ at $t=0$, defined in (\ref{Deltak}).

For illustration we evaluated the quantity  $\mu_0^2$ for $K=3$, using the
value $s_1=0.0033 \,{\rm fm }^{2}$ obtained recently in \cite{PoSc} and
several choices for  the higher derivatives $s_2$ and $s_3$. A few results are
listed in Table 1. The sets of values  for which $\mu_0^2$  is less
than 1 are consistent with the dispersive bounds (in particular, this is true
for $s_2=s_3=0$ and small deviations from these values, showing that 
Sirlin's  theorem is verified within the present formalism).  We give also
some pairs $(s_2,\,s_3)$ for which the quantity  $\mu_0^2$ is greater than
one, and therefore the inequality (\ref{consis1}) is violated. 
\begin{table}\label{tab2}
\begin{center}
\begin{tabular}{|c|c|c|}\hline &&\\  $s_2$ (${\rm fm}^4$)&$s_3$ (${\rm
fm}^6$)&$\mu_0^2$ \\&&\\ \hline 0.0&0.0&0.04\\ \hline
0.1&$-$0.1&0.69\\  \hline 
0.1&0.1&0.06\\  \hline 
$-$0.1&0.1&0.97\\  \hline 
$-$0.01&0.3&1.07\\  \hline 
0.05&$-$0.3&1.26\\ \hline
$-$0.1&0.5&4.71\\ 
\hline
\end{tabular}\end{center}
\caption{ The quantity $\mu_0^2$ defined in (\ref{minnorm1}) for 
$s_1=0.0033 \,{\rm fm }^{2}$ \cite{PoSc} and several values of the 
derivatives $s_2$ and $s_3$ of the Sirlin function. The values consistent with
QCD and analyticity correspond to $\mu_0^2\le 1$. } \end{table}
The results show an adequate  consistency between the requirements of
Sirlin's theorem on one hand, and QCD and analyticity  on the other. However,
in order to exploit the dispersive inequalities we  had to adopt a model for
the phase of the kaon form factors along the unphysical region. The
presence of the unphysical cut reduces therefore the  model independence of
the conclusions.

\section{Bounds on the Taylor coefficients of the pion form factor at zero
momentum}  In this Section we shall apply the dispersive formalism to the pion
electromagnetic form factor $F_\pi(t)$. We consider the Taylor expansion of
this function  around the origin $t=0$ \cite{GaLe}
\begin{equation}\label{taylor} F_{\pi}(t)=1+{1\over 6}\,r^2_\pi \,t +c \,t^2 +
d\, t^3 + .....\,, \end{equation} where   $r_\pi^2$ is the radius of the
charge distribution.  In
CHPT  the calculation of the Taylor coefficients requires the evaluation of 
higher pion loops, which introduce arbitrary renormalization constants
\cite{GaMe}, \cite{BiCo}.  One-loop CHPT \cite{GaLe} predicts  $c\approx
0.626\,{\rm GeV}^{-4},\, d\approx 2.30\, {\rm GeV}^{-6}$. At two - loop level
the coefficient $c$ can not be calculated as it depends on an arbitrary
renormalization constant, and  $d=4.1\, {\rm GeV}^{-6}$ \cite{BiCo}.   A fit
of the ALEPH data \cite{Aleph} on the hadronic $\tau$ decay rate with a
Gounaris-Sakurai formula \cite{GoSa} (equivalent  to the  Pad\'e version of
the one-loop CHPT)  gives  $c=3.72\,{\rm GeV}^{-4},\, d=9.80 \, {\rm
GeV}^{-6}$. Similar values  $c=3.9\,{\rm GeV}^{-4},\, d=9.70 \, {\rm
GeV}^{-6}$  were obtained in  \cite{Truo} by usual dispersion relations. Other
values proposed in the literature are: $c=4.1 \,{\rm GeV}^{-4}$ \cite{Dono},
and  $c=-7.5\,{\rm GeV}^{-4},\, d=62.5\, {\rm GeV}^{-6}$ 
\cite{DaGi}. 

In what follows we shall
prove that the dispersive formalism imposes nontrivial constraints on the
Taylor expansion (\ref{taylor}), especially when
combined with additional information about the form factor along a part of
the time like region. For the present purpose we   keep only the contribution
of the pion form factor in the unitarity inequality (\ref{unitelm}),
neglecting the positive terms due to the $K$ mesons. In this case the
inequality (\ref{l2norm}) becomes
 \begin{equation}\label{l2pi}  {1\over
2\pi} \int\limits_0^{2\pi}   \vert w_\pi(\zeta) F_\pi(\zeta)  \vert ^2 {\rm
d}\theta\leq 1\,, \quad\zeta={\rm exp}(i\theta)\,, \end{equation}  where the
outer function $w_\pi(z)$ is
\begin{equation}\label{wpi}  w_\pi(z)={(1-d_\pi)^2\over 16}\,\sqrt{{1\over
6\pi t_\pi \Pi'_{elm}(-Q^2)}}\, {(1+z)^2 \sqrt{1-z}\over (1-zd_\pi)^2}\,, 
\end{equation} with $d_\pi=d_1$ defined in (\ref{di}).   If we  use as an
alternative input the lower bound on the hadronic contribution to the muon
anomaly, the function $w_\pi(z)$  is
 \begin{equation}\label{wamupi} w_\pi (z)={1\over
16}\sqrt{{1\over 6\pi t_\pi a^{(h)}_\mu}}\,(1+z)^2 \sqrt{1-z}~w_{\cal
K}(z)\,, \end{equation}  where $w_{\cal K}(z)$ was defined below
(\ref{wamu}). 
 We consider now the function \begin{equation}\label{g_pi}  g(z)=w_\pi(z)
F_\pi(z) \,,\end{equation}  which is real analytic in the unit disk  $|z|<1$
and  can be expanded as \begin{equation}\label{g_piseries} 
g(z)=\sum\limits_{n=0}^\infty c_n z^n\,, \end{equation}  with real
coefficients $c_n=c_n^*$.  Then the condition (\ref{l2pi}) becomes
\begin{equation}\label{l2pi1}   \sum\limits_{n=0}^\infty  c_n^2 \leq 1\,.
\end{equation} 
A constraint on the first $N$  coefficients appearing in
the Taylor expansion (\ref{taylor}) is immediately obtained by  noticing that
(\ref{l2pi1}) implies \begin{equation}\label{l2piN}  
\sum\limits_{n=0}^N c_n^2 \leq 1\,.  \end{equation}  
The first $N$ coefficients $c_n$  can be
expressed  in a straightforward way in terms of the first $N$ Taylor
coefficients appearing in (\ref{taylor}) .  The explicit relations are obtained easily 
 using the definition
(\ref{g_pi}) and the conformal mapping (\ref{mapp}).  For instance,
$c_0=w_\pi(0)$, $c_1=w_\pi'(0)-2r_\pi^2t_\pi w_\pi(0)/3$ etc, where the
function $w_\pi(z)$ is defined in (\ref{wpi}) and the derivatives are with
respect to $z$.  If we use as input the lower bound on the muon anomaly, the
function $w_\pi(z)$ has the expression (\ref{wamupi}).

In particular, for $N=3$  and a fixed value of the charge radius
$r^2_\pi$ the inequality  (\ref{l2piN}) defines an allowed domain in the plane
of the coefficients $c$ and $d$ appearing in the Taylor expansion (\ref{taylor}). In
Fig.1, the interior of the larger ellipse is the domain obtained 
using the standard value  $r^2_\pi=0.42\, {\rm fm}^2$ and the  QCD
condition (\ref{condit}). This domain is very large, but one can
see a certain correlation among the values of the Taylor coefficients.
This feature becomes more stringent when higher derivatives of the
form factor are taken into account.  For comparison we indicated also in
Fig. 1 the allowed domain obtained from  the condition  (\ref{condit1}) on the
muon anomaly. This domain is  smaller, and in particular it excludes a pair
of values for $c$ and $d$ proposed recently in \cite{DaGi}.   
\begin{figure} \centerline{\epsfxsize=10cm\epsffile{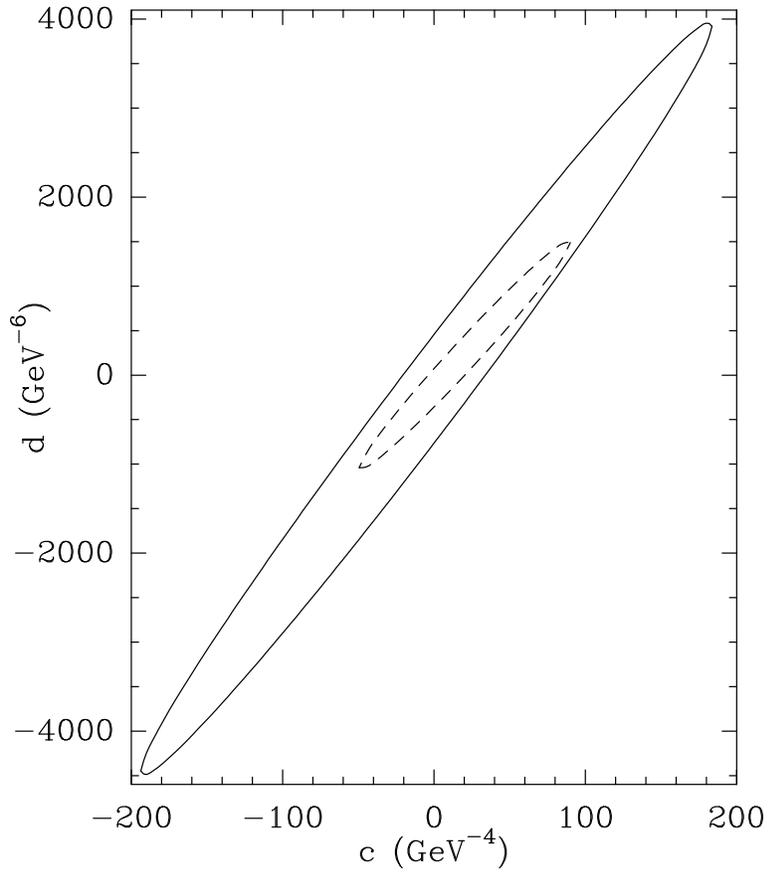}}
\caption{Allowed domains for the Taylor coefficients $c$ and $d$ of  the pion
electromagnetic form factor obtained from (\ref{l2piN}). The large ellipse is
obtained using as input the QCD expansion of the polarization function, the
small one using the lower bound on the muon  hadronic anomaly.}
\vspace{0.5 cm} \end{figure}  In the next subsections we shall improve the
above bounds by  implementing  informations about the phase and the modulus of
the pion form factor along a part of the unitarity cut.

\subsection{Improved bounds
using the phase of the form factor along a part of the cut} \label{ss1} 
According to Watson theorem \cite{Wats}, along the elastic region $t_\pi< t <
16 m_\pi^2$  the phase of the pion electromagnetic form factor coincides with
the phase $\delta_1^1(t)$ of the $L=1\,,\, I=1$  partial  wave amplitude of the
$\pi \pi$ scattering. This amplitude was calculated in the frame of CHPT up to
two loops \cite{BiCo}. The expansion upon chiral loops  is assumed to describe
correctly  the  phase up to $\approx 0.400\, {\rm GeV}$. On the other hand,
precise experimental data are available above $\approx0.600\, {\rm GeV}$
\cite{PiPi}. New planned experiments on $K_{l4}$ decay \cite{Loew},
\cite{Poca}, \cite{ChDy} will provide  accurate information about the low
energy $\pi\pi$ scattering amplitudes.

It is of interest to incorporate in the
dispersive formalism  for the pion form factor the additional information on
the phase.   Unlike the situation encountered in Section 3, where  the
knowledge of the phase allowed us to remove the unphysical cut below the
unitarity threshold,  in the present case we must implement the phase   along
a part of the unitarity cut. We treat the problem adapting   mathematical
techniques used  in \cite{Bour}-\cite{Capr2}.  We assume that
\begin{equation}\label{phase}
{\rm Arg}[F_\pi(t+i\epsilon)]=\delta_1^1(t)\,,\quad t_\pi\le t \le t_{in}\,,
\end{equation}
where $\delta_1^1 (t)$ is a known function and $t_{in}$ denotes the threshold of
the inelastic cut or a certain point up to which the phase is supposed to be
known.  As we shall discuss below, the technique can be easily adapted to
more general situations, for instance  when the phase is given along the range
$t_1\le t\le t_{in}$ with $t_1>t_\pi$, or along a region  consisting of two
disjoint parts, $t_\pi\le t\le t_1$ and  $t_2\le t\le t_{in}$, with $t_2>t_1$. 

We first express the condition (\ref{phase})
in the variable $z$ defined in (\ref{mapp}). To this end we  recall that the
branch point $t_\pi$ is mapped onto the point $z=-1$, and the upper edge of the
unitarity cut in the $t$ plane becomes the lower unit semicircle in the $z$
plane. We denote by $z_{in}=\exp (i\theta_{in})$ the image of the point
$t_{in}-i\epsilon$ in the $z$-plane , and let $\Gamma=\{\theta:
\theta_{in}<\theta<2\pi-\theta_{in}\}$. Then the  condition (\ref{phase})
becomes \begin{eqnarray}\label{phasez} \lim_{r\to 1}{\rm Arg}[F_\pi(r{\rm
e}^{i\theta})]&=&-\delta_1^1(\theta)\,,\quad
\theta_{in}\le \theta\le \pi\,,\nonumber\\ \lim_{r\to 1} {\rm Arg}[F_\pi(r{\rm
e}^{i\theta})]&=&\delta_1^1 (\theta)\,,\quad
\pi\le \theta\le 2\pi-\theta_{in}\,,\end{eqnarray}  where we denoted
$\delta_1^1(\theta)=\delta_1^1(t(\theta))$, with $t(\theta)=t_\pi+t_\pi {\rm
cotg}^2 (\theta/2)$ as  follows from  (\ref{mapp}).  We took into account the
fact that the form factor is an analytic function of real type, which means
that its phase is an odd function of $\theta$, {\it i.e.} it satisfies the
relation  $\delta_1^1 (\theta)=-\delta_1^1 (2\pi-\theta)$. It is useful to
note that the same property holds also for the imaginary part of the form
factor, while the real part and the modulus are even functions of $\theta$. 

In what follows we shall derive the allowed domain for
the Taylor coefficients of the expansion (\ref{taylor}), taking into account
the inequality (\ref{l2pi}) provided by the dispersive formalism, and the
additional relations (\ref{phasez}). As usual, the information about the
phase is implemented by means of an  Omn\`es function \cite{Omne}. We can use
the definition given in (\ref{O_i}), or equivalently we 
define \begin{equation}\label{omnes} {\cal O}_\pi(z)={\rm
exp}\left[{i\over \pi}\int\limits_0^{2\pi}{\rm d}\theta {\bar \delta_1^1
(\theta)\over 1-z{\rm e}^{-i\theta}}\right]\,, \end{equation} where  $\bar
\delta_1^1 (\theta)$ is a Lipschitz continuous function such that
$\bar\delta_1^1(\theta)=-\delta_1^1(\theta)\,$ for $\,\theta_{in}<\theta
<\pi\,,$$\,\,\bar\delta_1^1(\theta)=\delta_1^1 (\theta)\,$ for $\,\pi<\theta
<\pi+\theta_{in}$,   $\bar
\delta_1^1 (\theta)$  being arbitrary outside $\Gamma$. We shall see below that
the results are independent on the choice of $\bar \delta_1^1(\theta)$ outside
$\Gamma$.  Using the Plemelj-Privalov relation \cite{Mush} 
\begin{equation}\label{plemelj}  \lim\limits_{r\to 1}{1\over
\pi}\int\limits_0^{2\pi}{\rm d}\theta' {F(\theta')\over 1-r{\rm
e}^{i(\theta-\theta')}}=F(\theta)+ {1\over \pi}\int\limits_0^{2\pi}{\rm
d}\theta' {F (\theta')\over 1-{\rm e}^{i(\theta-\theta')}}\,, \end{equation}
where the last integral denotes the Principal Value,
one can show that along the interval  $\Gamma$ the phase of the function
${\cal O}_\pi$ coincides with the phase of the form factor.
Therefore, by multiplying $F_\pi(z)$ with  $[O_\pi(z)]^{-1}$ the phases
compensate each other and the product is real along  the elastic part of the
cut. In the variable $z$ this condition has the form
\begin{equation}\label{imag} {\rm Im}\lim_{r\to 1}\left[{1\over {\cal O}_\pi
(r{\rm e}^{i\theta})}\,F_\pi(r{\rm e}^{i\theta})\right] =0\,,\quad \theta\in
\Gamma\,. \end{equation}  By recalling the definition (\ref{g_pi}) of the
function $g(z)$ and its power expansion (\ref{g_piseries}),
we write the condition (\ref{imag}) in the form
\begin{eqnarray}\label{imag1} {\rm Im}\lim_{r\to 1}\left[{1\over w_\pi(r{\rm
e}^{i\theta})  {\cal O}_\pi(r{\rm e}^{i\theta})}\,  g(r{\rm
e}^{i\theta})\right]= \nonumber\\\sum\limits_{n=0}^\infty c_n\, {\rm
Im}\lim_{r\to 1}\left[[W(r{\rm e}^{i\theta})]^{-1}\, r^n\,{\rm e}^{i
n\theta}\right] =0\,,\quad \theta\in \Gamma\,, \end{eqnarray}  where $W(\zeta)$
is  defined as  \begin{equation}\label{W}  W(\zeta)=w_\pi(\zeta) {\cal
O}_\pi(\zeta) \,,\quad \zeta={\rm e}^{i\theta}\,. \end{equation} 
The allowed domain of the
Taylor coefficients of the pion form factor, which  satisfy the conditions
(\ref{l2pi}) and (\ref{imag}) can be found, like in Section 3,  by means of
 an optimization problem.  We first recall that the conditions
(\ref{l2pi}) and (\ref{imag}) were written in the equivalent forms
(\ref{l2pi1}) and  (\ref{imag1}) respectively, in terms of the Taylor
coefficients $c_n$.  We consider then the quantity
\begin{equation}\label{mu_0pi} \mu_0^2= \min_{c_n} \sum\limits_{n=0}^\infty 
c_n^2\,, \end{equation} where the minimization  is with respect to the
coefficients $c_n$ which satisfy the condition (\ref{imag1}).
As in Section 3 one can show that the inequality
(\ref{consis}) is a
necessary and sufficient condition for the consistency of the relations  
(\ref{l2pi1}) and  (\ref{imag1}).

We solve the constrained minimization problem (\ref{mu_0pi})
with the generalized Lagrange theory of optimization, based on
Hahn-Banach theorem \cite{Luen}. The Lagrangean is
\begin{equation}\label{L_pi}   {\cal L}=\sum\limits_{n=0}^\infty  c_n^2
+{2\over \pi}\sum\limits_{n=0}^\infty c_n \lim_{r\to 1}\int\limits_\Gamma
\lambda(\theta)\, |W(\theta)|\, {\rm Im}\left[[ W(\theta)]^{-1}\, r^n{\rm
e}^{i n\theta}\right]{\rm d}\theta \,,\end{equation}  where the function 
$\lambda (\theta)$ is a generalized Lagrange multiplyer \cite{Luen}. We can
assume without losing generality that it is an odd function, since its product
 with another odd function (the imaginary part of an analytic function of real
type) is integrated along a symmetric interval $\Gamma$. We denoted for
simplicity $W(\theta)=W(\exp(i\theta))$. The numerical factor in front of the
integral  in (\ref{L_pi}) and the  modulus $|W(\theta)|$  inside the integral
were introduced  explicitely for convenience. 

We assume now that the first $N$ coefficients
$c_n$ have prescribed values, and perform the minimization of the Lagrangean
with respect to the remaining coefficients $c_n$, $n\ge N+1$. The minimum
condition \begin{equation}\label{minimpi} {\partial{\cal L}\over \partial
c_{n}}=0\,,\quad n\ge N+1\end{equation} 
gives the optimal coefficients
 \begin{equation}\label{c_npi}
c_n=-{1\over \pi} \lim_{r\to 1}\int\limits_\Gamma \lambda(\theta)\,
|W(\theta)|\, {\rm Im}\left[[ W(\theta)]^{-1}\, r^n\,{\rm e}^{i
n\theta}\right]{\rm d}\theta \,,\quad n\ge N+1\,.\end{equation} 
This expression can be written
equivalently as  \begin{equation}\label{c_npi1}
c_n=-{i\over \pi} \lim_{r\to 1}\int\limits_\Gamma \lambda(\theta)\,
|W(\theta)|\, [ W^*(\theta)]^{-1}\,r^n\, {\rm e}^{-i
n\theta}{\rm d}\theta \,,\quad n\ge N+1\,,\end{equation} 
where we added a term which has a vanishing contribution by parity arguments. 

In order to find the Lagrange multiplier $\lambda(\theta)$ we introduce the
optimal coefficients $c_n$ in the
constraint (\ref{imag1}). Recalling that the
first $N$ coefficients  $c_n$ are assumed to have prescribed values, we
write the condition (\ref{imag1}) in the form \begin{equation}\label{inteq0}
\sum\limits_{n=0}^N c_n\, {\rm Im}\left[{{\rm e}^{i n\theta} \over
W(\theta)}\, \right] -{\rm Im}\lim_{r\to 1}\left[{i\over \pi}{1\over
W(\theta)} \int\limits_\Gamma \lambda(\theta')\, {|W(\theta')|\over
W^*(\theta')} \,{{\rm e}^{i(N+1)(\theta-\theta')}\over 1-r {\rm
e}^{i(\theta-\theta')}}\,{\rm d}\theta'\right]=0\,,\,\,\theta\in \Gamma\,.
\end{equation}  It is useful to write
 \begin{equation}\label{phaseW}
W(\theta)= |W(\theta)|{\rm e}^{i\Phi(\theta)}\,,
\end{equation}
where from (\ref{W}) it follows that  
\begin{equation}\label{phaseW1} \Phi(\theta)=\phi(\theta)+
\bar\delta_1^1(\theta)\,.\end{equation} Here $\phi(\theta)$ is the phase of
the outer function $w_\pi(z)$,  and
$\bar\delta_1^1(\theta)$  is  connected to the  phase of the pion
form factor as explained below (\ref{omnes}). 
By inserting (\ref{phaseW})
 in the relation (\ref{inteq0}) and
applying the Plemelj-Privalov relation (\ref{plemelj}) we obtain  after a
straightforward calculation  the equation \begin{eqnarray}\label{inteq1}
&&\sum\limits_{n=0}^N c_n\, \sin[  n\theta- \Phi(\theta)]-\lambda(\theta)
\nonumber\\ &&+{1\over 2\pi} \int\limits_\Gamma \lambda(\theta')\,
{\sin\left[\left(N+{1\over 2}\right)(\theta-\theta')-\Phi(\theta)
+\Phi(\theta')\right] \over \sin \left[{\theta-\theta'\over 2}\right]}\,{\rm
d}\theta'=0\,,\,\,\theta\in \Gamma\,, \end{eqnarray} where the last integral
is  defined as the Principal Value.  We obtained  therefore a singular 
integral equation for the Lagrange multiplier $\lambda(\theta)$. If the
phase $\Phi(\theta)$ is Lipschitz continuous, the equation is of Fredholm type
and can be solved by standard techniques.  After solving this
equation, the minimal norm $\mu_0^2$ can be computed by inserting in
(\ref{mu_0pi}) the prescribed values $c_n$ for $n\le N$ and the optimal values
(\ref{c_npi1}) of $c_n$ for $n\ge N+1$ . Taking into account
the fact that $c_n=c_n^*$ we obtain  \begin{equation}\label{mu_0pi1}  
\mu_0^2= \sum\limits_{n=0}^N c_n^2+   \lim_{r\to 1} {1\over \pi^2} \lim_{r\to
1}\int\limits_\Gamma {\rm d}\theta \int\limits_\Gamma {\rm d}\theta' 
\lambda(\theta)\,\lambda(\theta')\,   {|W(\theta)|\over W^*(\theta)}\, 
{|W(\theta')|\over W^*(\theta')}\,{{\rm   e}^{i(N+1)(\theta-\theta')}\over 1-r
{\rm e}^{i(\theta-\theta')}}\,. \end{equation} By applying the Plemelj-Privalov
relation (\ref{plemelj}) and using  the integral equation
(\ref{inteq1}) satisfied by the function $\lambda (\theta)$, we arrive finally
at the expression \begin{equation}\label{mu_0pi2}   \mu_0^2=
\sum\limits_{n=0}^N c_n^2+\sum\limits_{n=0}^N {c_n \over
\pi}\,\int\limits_\Gamma  \lambda(\theta) \,\sin[n\theta-\Phi(\theta)]{\rm
d}\theta\,. \end{equation} Using the fact that $\lambda (\theta)$ and $\Phi
(\theta)$ are odd functions, the integral equation (\ref{inteq1}) can be
written in the form  \begin{eqnarray}\label{inteq2}
&&\sum\limits_{n=0}^N c_n\, \sin[  n\theta- \Phi(\theta)]-\lambda(\theta)
\nonumber\\
 &&+{1\over 2\pi} \int\limits_{\theta_{in}}^\pi \lambda(\theta')\,
\left[{\sin\left[\left(N+{1\over 2}\right)(\theta-\theta')-\Phi(\theta)
+\Phi(\theta')\right] \over \sin \left[{\theta-\theta'\over
2}\right]}\right.\nonumber\\ &&-\left. {\sin\left[\left(N+{1\over
2}\right)(\theta+\theta'-2\pi)-\Phi(\theta) -\Phi(\theta')\right] \over \sin
\left[{\theta+\theta'\over 2}-\pi\right]}\right]\, 
{\rm d}\theta'=0\,,\quad \theta_{in}\leq\theta\leq \pi\,,\end{eqnarray} 
where
\begin{equation}\label{phaseW2} \Phi(\theta)=\phi(\theta)-
\delta_1^1(\theta)\,,\quad \theta_{in}\leq\theta\leq \pi\,. \end{equation}
Moreover, the expression (\ref{mu_0pi2}) of $\mu_0^2$ can be written as 
 \begin{equation}\label{mu_0pi3} 
 \mu_0^2= \sum\limits_{n=0}^N c_n^2+\sum\limits_{n=0}^N \,{2 \,c_n \over \pi}
\,\int\limits_{\theta_{in}}^\pi \lambda(\theta)
\,\sin[n\theta-\Phi(\theta)]{\rm d}\theta\,. \end{equation} We notice
that the first term  in  (\ref{mu_0pi3}) represents the unconstrained minimum,
obtained from (\ref{l2piN}) if the coefficients $c_n$ for $n\ge N+1$ were
free. The second term in the  expression of $\mu_0^2$ is positive and
represents the improvement brought by the knowledge of the phase of the form
factor along the region $t_\pi<t<t_{in}$.

It is important to emphasize that only the values
of $\Phi(\theta)$  along the interval  $\Gamma$ are required in the integral
equation. Also, the modulus of the outer function  does not appear (it was
absorbed  in the definition of the Lagrange multiplier $\lambda (\theta)$),
which means that  the results are independent of the choice of the function
$\bar\delta_1^1$ outside the interval $\Gamma$.  

The equations (\ref{inteq2}) - (\ref{mu_0pi3}) provide a simple
numerical procedure for finding the allowed domain of the coefficients of the
Taylor expansion (\ref{taylor}): one starts with a set of
values for the first $N$ Taylor coefficients. Using  the relations
(\ref{g_pi}) and (\ref{g_piseries}) one finds the corresponding
coefficients $c_n$,  which enter  the integral equation
(\ref{inteq2}). The solution $\lambda (\theta)$ of
this equation is then used in (\ref{mu_0pi3}) to evaluate the quantity
$\mu_0^2$.  Recalling that the allowed domain of the Taylor
coefficients is described by the inequality (\ref{consis}), the 
values taken as input are accepted or rejected if
 $\mu_0^2$ is less or greater than unity, respectively. 

From the above derivation the generalization to the case where the region
$\Gamma$ consists of two disjoint subintervals $\Gamma_1$ and $\Gamma_2$  is
straightforward. This case is of interest when using the phase given by CHPT
along $\Gamma_1=\{t: \,t_\pi<t< (0.400\, {\rm GeV})^2\}$ and the experimental
data on    $\Gamma_2=\{t: \,(0.600\, {\rm GeV})^2<t<t_{in}\}$. The resulting
equations have the same form, the Lagrange multiplier being defined on
$\Gamma_1 \cup \Gamma_2$.

 To illustrate how the bounds on the Taylor coefficients
$c$ and $d$ are improved by the knowledge of the phase, we took 
the expression \cite{GuPi}  \begin{equation}\label{phaseunit} 
\delta_1^1(t)={\rm arctg}\left[\,{m_\rho \Gamma_\rho(t)\over
m_\rho^2-t}\,\right]\,, \end{equation} where $\Gamma_\rho(t)$ was defined in
(\ref{gamaro}).
 At low
energies above the threshold, this phase coincides with 
the  one loop CHPT   expression
\cite{GaLe} \begin{equation}\label{phasechpt} \delta_1^1(t)= {t\over 96 \pi
f_\pi^2}\,\left(1-{4m_\pi^2\over t}\right)^{3/2}\,,  \end{equation} 
while for $t\ge \,(0.500 {\rm GeV})^2$ it is in very good agreement with the
experimental data \cite{PiPi}.
  We assumed that 
the phase of the pion form factor coincides with (\ref{phaseunit}) along the
region  $t_\pi<t<t_{in}$, with  $t_{in}= 0.8\,{\rm GeV}^2$, which corresponds
to $\theta_{in}=0.6321$.
As input for the bounds we use   
the OPE expansion of the polarization amplitude, which led to the
domain represented by the interior of the large ellipse in  Fig.1. In this
case  the outer function $w_\pi(z)$ is given in 
(\ref{wpi}) and the function $\Phi (\theta)$ defined in (\ref{phaseW2}) has
the expression  \begin{equation}\label{phaseW3} \Phi(\theta)={5\theta-\pi
\over 4}+{\rm arctg}\left[{2d_1\sin \theta -d_1^2\sin 2\theta\over
1-d_1\cos\theta+d_1^2 \cos 2\theta}\right] -\delta_1^1(\theta)\,,\quad
\theta_{in}\leq \theta\leq \pi\,,\end{equation}  with $d_1$ defined in
(\ref{di}).  
\begin{figure} \centerline{\epsfxsize=10cm\epsffile{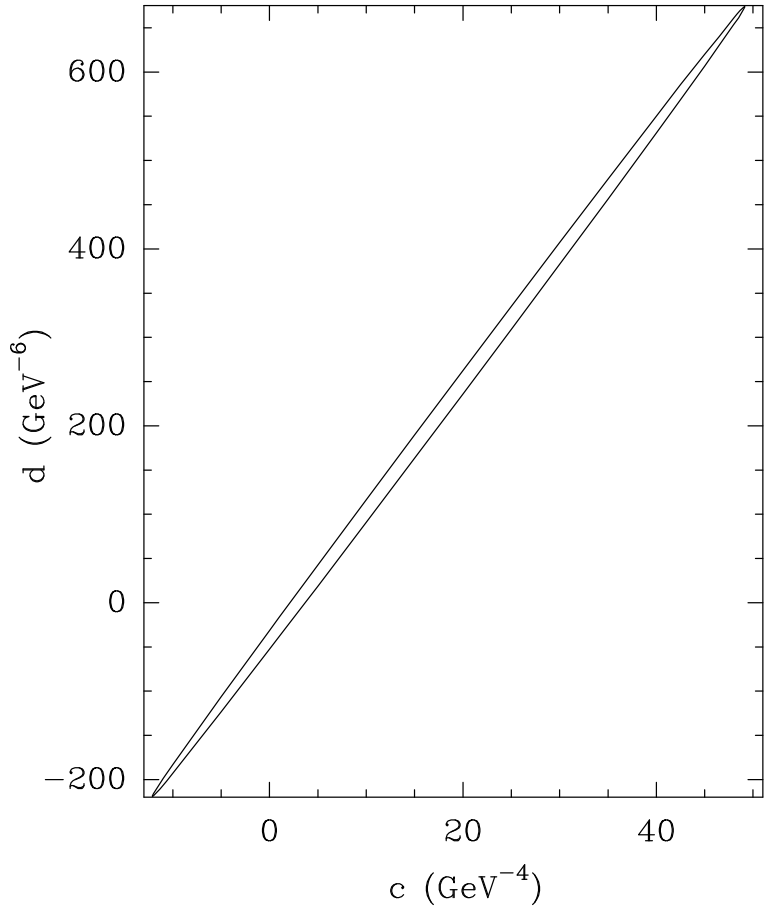}}
\caption{Improved domain for the Taylor coefficients $c$ and $d$ of the pion
electromagnetic form factor,
 using the phase along a part of the boundary.} \vspace{0.5 cm} \end{figure}
In Fig. 2 the
interior of the ellipse is the
allowed domain in the plane  $(c, d)$, for $r_\pi^2=0.42 \,{\rm fm}^2$.  
Compared with the large ellipse in Fig.1 one can see the considerable
improvement brought by the knowledge of the phase along a part of the
unitarity cut. A more realistic calculation    using CHPT along an interval
$\Gamma_1$ and the experimental data on $\delta_1^1(t)$ along another interval
$\Gamma_2$ is of interest and will provide precise model independent bounds on
the Taylor coefficients.

\subsection{Improved bounds using information about the phase and the
modulus  along a part of the cut}\label{ss2}
The modulus of the pion form factor along a part of the cut is known
experimentally from the rate of $e^+e^-$ annihilation into pions \cite{BaBi}
and the hadronic $\tau$ decay \cite{Aleph}. 
 High precision data are available especially in the range 
$0.3\, {\rm GeV}^2\,<\,t\,<\,0.9\, {\rm GeV}^2$.  In  this subsection we 
derive the allowed domain of the Taylor coefficients by including some
information about the modulus  of the pion  form factor along a part of the
cut. 

It is a known mathematical fact that if the modulus and the phase
of an analytic function are known exactly along  a part of the boundary, then
the function is in principle uniquely determined. Explicit formulas for
recapturing an analytic function  belonging to a certain class, from its
restriction  along    a part of the boundary are available \cite{KrNu},
\cite{RoRo} (see also \cite{Capr1} for explicit expressions and further
references). These expressions are however very unstable numerically,
reflecting the fact that the analytic continuation is an ill-posed problem
\cite{CiPo}. More exactly, the formulas give the correct analytic continuation
if the input values are known with infinite accuracy, but they lead to
arbitrary predictions if these values are affected by errors.  

In what
follows we shall show that even a nonoptimal  use of the input information
about the phase and the modulus  leads to a considerable improvement of the
bounds on the Taylor coefficients.  Unlike the  case considered in the previous
subsection, where the phase could be given along  several disjoint intervals,
the  method  described below requires the knowledge of the phase along the
whole range $t_\pi\le t\le t_{in}$.  We start by writing the inequality
(\ref{condelm}) in the form \begin{equation}\label{condlb}   {1\over
48\pi^2}\int\limits_{t_{in}}^\infty{{\rm d}t \over (t+Q^2)^2} 
\left(1-{4m_\pi^2\over t}\right)^{3/2}   \vert F_{\pi}(t) \vert ^2 \leq {\cal
M}(Q^2)\,, \end{equation}  where \begin{equation}\label{M0}   {\cal
M}(Q^2) =\Pi'_{elm}(-Q^2)- {1\over 48\pi^2}\int\limits_{t_\pi}^{t_{in}}{{\rm
d}t \over (t+Q^2)^2}  \left(1-{4m_\pi^2\over t}\right)^{3/2}  |F_\pi(t)|^2 \,.
\end{equation}   Assuming that the modulus $|F_\pi(t)|$ is known along
$t_\pi\le t\le t_{in}$, the quantity ${\cal M}(Q^2)$ can be evaluated
numerically. To be conservative, one can use a lower bound on the modulus,
which does not spoil the inequality (\ref{condlb}) and leads to a larger
allowed domain  for the Taylor coefficients. 

 In order to incorporate the information about the phase we use the
 Omn\`es function defined in (\ref{omnes}), which
can be written equivalently in the $t$ variable as
\begin{equation}\label{O_pi}  O_{\pi}(t)={\rm exp}\left[{t\over
\pi}\int\limits_{t_\pi}^{\infty}{\bar \delta_1^1(t') \over t'(t'-t)}{\rm
d}t'\right]\,.\end{equation} The  properties of the function $\bar
\delta_1^1(t)$ were discussed below  (\ref{omnes}), and we use it in order to
remove the part of the cut below $t=t_{in}$. More precisely, writing as in
(\ref{f_i}) \begin{equation}\label{f_pi} F_\pi(t)=f_\pi(t) O_\pi(t)\,, 
\end{equation} we see that the function $f_\pi(t)$ is real on the real axis
below $t_{in}$, since the phase of the form factor along $t_\pi\le t\le
t_{in}$ is compensated by  the phase of $O_\pi$. Therefore $f_\pi(t)$ is
analytic in the $t$-plane cut for $t>t_{in}$, and satisfies on the cut the
condition  \begin{equation}\label{condlb1}   {1\over
48\pi^2}\int\limits_{t_{in}}^\infty{{\rm d}t \over (t+Q^2)^2} 
\left(1-{4m_\pi^2\over t}\right)^{3/2}  \vert f_{\pi}(t) \vert ^2    \vert
O_{\pi}(t) \vert ^2 \leq {\cal M}(Q^2)\,, \end{equation}  which follows from
(\ref{condlb}). But this condition can be simply exploited  with the standard
techniques described in subsection \ref{s3}. We first map the $t$ plane  cut
for $t>t_{in}$ onto the unit disk in the $z$-plane, using the conformal
mapping  \begin{equation}\label{tildemapp} z(t)={\sqrt
{t_{in}-t}-\sqrt{t_{in}}\over \sqrt {t_{in}-t}+\sqrt{t_{in}}}\,.
\end{equation}  We define as in (\ref{omega1}) the outer function  
\begin{equation}\label{omegapi}  \omega_\pi(t) ={\rm exp}
\left[{\sqrt{t_{in}-t}\over\pi}\int\limits_{t_{in}}^\infty  {\ln \vert
O_\pi(t')\vert \over \sqrt{t'-t_{in}} (t'-t)}{\rm d}t'\right]\,, \end{equation}
such that $|\omega_\pi(t)|=|O_\pi(t)|\,$ for $ t>t_{in}$,
and the outer function similar to (\ref{wpi})
\begin{equation}\label{tildewpi} \tilde w_\pi(z)={(1-\tilde d_\pi)^2\over
16}\,\sqrt{{1\over 6\pi t_\pi {\cal M}(Q^2)}}\, {(1+z)^2 \sqrt{1-z}\over
(1-z\tilde d_\pi)^2}\,, \,, \end{equation} with 
\begin{equation}\label{tildedpi}
\tilde d_\pi={\sqrt{t_{in}+Q^2}-\sqrt{t_{in}}
\over\sqrt{t_{in}+Q^2}+\sqrt{t_{in}}}\,.
\end{equation}
 By introducing a new function
$\tilde g(z)$, defined as \begin{equation}\label{tildeg_pi} \tilde g(z) =
\tilde w_\pi(z) \omega_\pi(z)
 f_\pi(z)\,=\,
\tilde w_\pi(z) \omega_\pi(z)
[O_\pi(z)]^{-1} F_\pi(z)\,,\end{equation} the
inequality (\ref{condlb1}) takes the canonical form
\begin{equation}\label{l2normpi3} {1\over 2\pi}
\int\limits_0^{2\pi}  \vert \tilde g (\zeta) \vert ^2 {\rm
d}\theta\leq 1\,,\quad \zeta={\rm exp}(i\theta)\,. \end{equation}
 By
construction the function $\tilde g(z)$ is analytic in the disk $|z|<1$ and can
be expanded  as   \begin{equation}\label{tildeg_piser}
\tilde g(z)=\sum\limits_{n=0}^\infty \tilde c_{n} z^n\,, \end{equation}
where the coefficients $\tilde c_n$ satisfy the inequality
 \begin{equation}\label{l2pi3} 
 \sum\limits_{n=0}^\infty  \tilde c_n^2 \leq 1\,,
\end{equation} 
which follows from (\ref{l2normpi3}).
\begin{figure} \centerline{\epsfxsize=10cm\epsffile{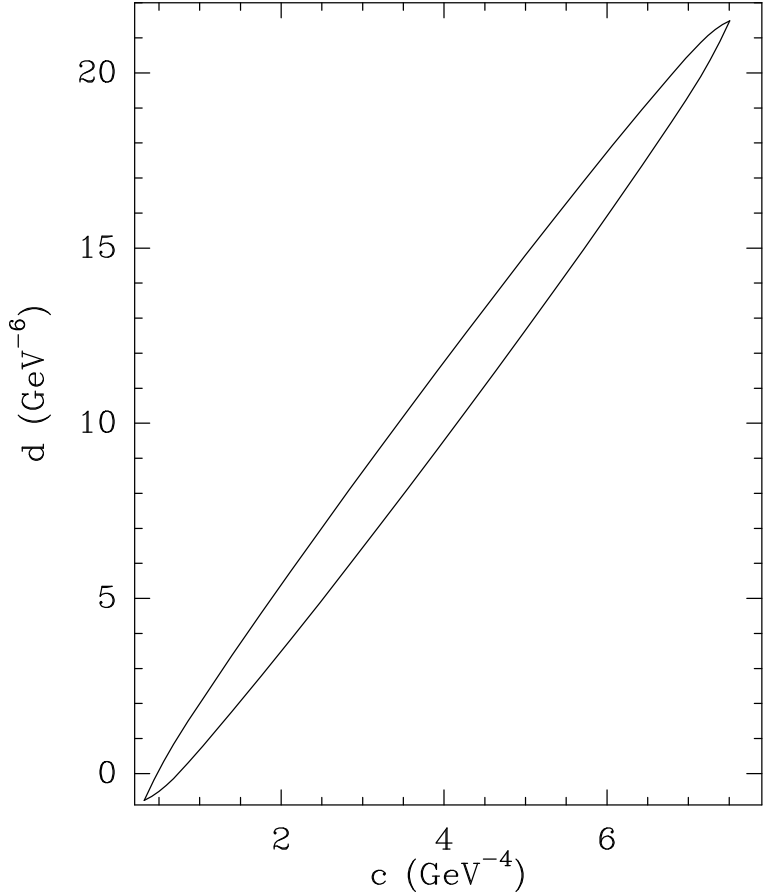}}
\caption{Improved domain for the Taylor coefficients $c$ and $d$ of the pion
electromagnetic form factor, 
using the phase and the modulus along a part of the boundary.} \vspace{0.5 cm}
\end{figure}
The first $N$  coefficients $\tilde c_n$ are connected in a
straightfoward way through (\ref{tildeg_pi}) to the first $N$ Taylor
coefficients of the form factor $F_\pi(t)$ at $t=0$. Therefore, the inequality
 \begin{equation}\label{l2pi3N} 
 \sum\limits_{n=0}^N \tilde c_n^2 \leq 1\,,
\end{equation} 
which follows from (\ref{l2pi3}), defines an allowed domain in the plane of
these Taylor coefficients. In principle this domain can be further
reduced: indeed,  in addition to the boundary condition  (\ref{condlb1}),  the
function $f_\pi$ defined in (\ref{f_pi}) has  known error-affected values
along the region $t_\pi\le t\le t_{in}$, where it is holomorphic. The
inclusion of this additional information is not trivial, and we will not treat
this problem here, using the simple inequality (\ref{l2pi3N}). In
particular, for $N=3$ and $r_\pi^2$ fixed, this inequality gives the allowed
domain of the parameters $c$ and $d$ of the expansion (\ref{taylor}). 
 
In a numerical application we used the same phase $\delta_1^1(t)$ as in the
preceeding subsection and the modulus $|F(t)|$ from  the
ALEPH data \cite{Aleph}. Using $Q^2=2 \,{\rm GeV}^2$  and
$t_{in}=0.8\, {\rm GeV}^2$,  we obtain  for the quantity ${\cal M}(Q^2)$
defined in (\ref{M0}) the estimate ${\cal M}(Q^2)=0.00689 {\rm GeV}^{-2}$. The
result is not sensitive to the experimental uncertainties above the threshold
$t_\pi$, due to the phase space factor in the unitarity integral appearing in
(\ref{M0}). Reducing the value of $|F(t)|$ along $t_\pi\le t\le t_{in}$ by $5
\%$ leads to an increase of  ${\cal M}(Q^2)$ by about  $3\%$ and to  bounds on
the coefficients $c$ and $d$ weaker by about $2\%$. The interior of the ellipse
shown in Fig.3 indicates the allowed domain obtained in this conservative
situation. 

 The  combined information on the
phase and the modulus is seen to restrict in an impressive way the values of
the Taylor coefficients  of the pion form factor. The allowed range of $c$ is
$(0. 25\,,\,7.57)$, and for each $c$ the range of the values of $d$ is very
narrow. For instance, for $c=3.85$ the parameter $d$ is restricted to the
interval $(9.02\,,\,11.30)$. The values $c=-7.5\,{\rm GeV}^{-4},\, d=62.5\,
{\rm GeV}^{-6}$  adopted in  \cite{DaGi}  are outside the allowed domain. We
recall that these strong bounds are obtained  with no  specific
parametrization and without any assumption about the high energy behaviour of
the form factor.
\section{Conclusions} In the present paper we  investigated the  form factors
of the light pseudoscalar mesons in a dispersive formalism which  uses as
input the OPE expansion of the QCD polarization functions,  combined with
hadronic unitarity and analyticity. We derived constraints on the size and
shape of the form factors, which are of interest in particular for testing the
low energy predictions of CHPT. Generalizations of the original mathematical
techniques, suitable for including additional  informations about the form
factors were developed. 

In Section 3 we performed a test of Sirlin's theorem \cite{Sirl}, 
\cite{PoSc},  which requires the vanishing of a certain combination of form
factors which is free of arbitrary renormalization constants in CHPT at $t\ne
0$. To this end we treated simultaneously in the dispersive formalism all the
electroweak form factors of the $\pi$ and $K$ mesons. A
difficulty is  the  large unphysical cut of the kaon electromagnetic form
factors, which spoils the model independence of the results.  This is in
contrast with the case of heavy mesons, where the weak form factors of the
ground state and of the excited ones have very close  branch points and the
unphysical cuts are well approximated by a few number of poles. In that case
the simultaneous treatment of several form factors related among them by heavy
quark symmetry  led to strong  model independent bounds near zero recoil
\cite{Capr}, \cite{BoGr1}, \cite{CaLe}. In the present paper we adopted 
for the phase of the kaon form factors along the unphysical cut a model
inspired from the resonance chiral effective theory, which turned out to be
consistent with QCD and Sirlin's theorem.  This
subsection has however mainly a pedagogical character, showing how to combine
in an optimal way the dispersive formalism  with the Omn\`es
functions and the technique  of Lagrange multipliers for the constraints
at interior points. 

The main results are presented in Section 4, where we derived  model
independent constraints on the Taylor coefficients of the pion
electromagnetic form factor.  We showed how to implement in the dispersive
formalism  the knowledge of the phase of the form factor along a
part of the unitarity cut, even when this part consists of several disjoint
intervals. To this end we applied the Lagrange theory for functional
optimization, which led to an integral equation for the generalized Lagrange
multiplier.  The knowledge of the phase  considerably improves
the simple bounds  on the higher Taylor coefficients 
yielded by the dispersive formalism. In the present paper we illustrated
this statement using a realistic model of the phase \cite{GuPi}, which
reproduces the one-loop CHPT expression \cite{GaLe} below $0.400 \,{\rm GeV}$
and the present experimental data \cite{PiPi}  above $0.500 \,{\rm GeV}$.  The
accurate data  provided on the phase $\delta_1^1$ by the future
$K_{l4}$ experiments \cite{Loew},\cite{Poca} will be a precious input to the
formalism.  In Section 4 we obtained also  improved constraints on the Taylor
coefficients if both the phase and the modulus of the form factor are given
along a part of the cut, even if this information is used in an nonoptimal
way. The  interior of the ellipse shown in Fig. 3 represents the allowed
domain of the coefficients $c$ and $d$ entering the Taylor expansion
(\ref{taylor}), obtained with the phase  discussed above and the modulus from
the hadronic $\tau$ decay rate \cite{Aleph} along $t_\pi\le t\le 0.8 \,{\rm
GeV}^2$. Once this input information is adopted, the bounds are model
independent, since they are not based on specific parametrizations and are
free of any assumption about the high energy behaviour of the form factors. 
It is of interest to apply the techniques described in this paper to the
higher Taylor coefficients. The results obtained so far suggest that strong
correlations among these coefficients are expected, leading to restrictions on
the arbitrary renormalization constants of CHPT.

\vskip0.3cm
\noindent
  {\bf Acknowledgements:} I am pleased to thank   Prof. H. Leutwyler 
 for very useful discussions and the kind hospitality at the
Institute of Theoretical Physics of the University of Berne, and Dr. C.
Bourrely for many suggestions and a careful reading of the manuscript.  The
support of the Swiss National Science Foundation in the program CSR CEEC/ NIS,
Contract No 7 IP 051219 is gratefully acknowledged. 
 \end{document}